\documentclass[UTF8]{revtex4-1}
\usepackage[utf8]{inputenc}
\usepackage{graphicx}% Include figure files
\usepackage{dcolumn}% Align table columns on decimal point
\usepackage{bm}% bold math
\usepackage{amsmath}
\usepackage{amssymb}
\usepackage{latexsym}
\usepackage{epsfig}
\usepackage{amsbsy}
\usepackage{array}
\usepackage{setspace}
\usepackage{mathrsfs}
\usepackage{geometry}
\usepackage{caption}
\usepackage{multirow}

\begin{document}

\thispagestyle{empty}
	
\title{Nonrelativistic expansions of Dirac-Coulomb energy and first order relativistic Breit potential corrections for light Atomic and molecular systems}% Force line breaks with \\
\author{Wanping Zhou  \footnote{Email: zhouwp@hgnu.edu.cn}}
\affiliation{School of Physics and Telecommunications, Huanggang Normal University, Huanggang, China, 438000}
\author{Sanjiang Yang \footnote{Email: sjyang@hbnu.edu.cn}}
\affiliation{College of Physics and Electronic Science, Hubei Normal University, Huangshi, China,  435002}
\author{Chen Tong \footnote{Email: 2018202020011@whu.edu.cn}}
\affiliation{School of Physics and Electrical Engineering, Zhengzhou Normal University, Zhengzhou, China,450044}

\begin{abstract}
The nonrelativistic expansion method is applied to the Dirac-Coulomb energy and first order relativistic Breit potential corrections for light atomic and molecular systems to derive high-order effective energy correction operators for few-body systems. It is found that effective operators at the $m\alpha^8$ order contain incompletely separable divergences. These divergences must be canceled jointly with divergences arising from second- and third-order perturbations of lower-order energy correction operators, and such cancellation can be accomplished automatically by the nonrelativistic numerical expansion approach. Using Gaussian basis sets, we numerically compute contributions to the Dirac-Coulomb energy and relativistic Breit potential corrections up to order $m\alpha^8$ for the ground states of the hydrogen atom, hydrogen molecular ion, helium atom, and hydrogen molecule. These results confirm the reliability of the present method and enable its extension to more complex systems.
\end{abstract}

%Use

\maketitle

\newpage

 \section{INTRODUCTION}

The most powerful approach to calculating high-order energy corrections in nonrelativistic atoms and molecules is the effective Hamiltonian method developed by Pachucki et al\citep{Annphy.226.1,PhysRevA.71.012503,JPhysB.31.5123,PhysRevA.72.062102,PhysRevA.100.042510,PhysRevA.103.042809}. This approach has achieved great success over the past few decades. For various high-order energy corrections, including relativistic kinetic energy corrections, photon electron-exchange corrections, electron-nucleus recoil corrections, loop diagram corrections, and mixed corrections from multiple physical effects, the effective Hamiltonian framework typically adopts the Foldy–Wouthuysen (FW) transformed Hamiltonian to evaluate virtual photon contributions in the low-energy regime, while high-energy contributions are computed via QED perturbation theory.

As the perturbation order increases, the number of terms required in the effective Hamiltonian grows rapidly, as can be clearly seen from the high-order expansion of the Foldy–Wouthuysen (FW) transformed Hamiltonian \citep{JPhyA.56.4,PhysRevA.100.012513}
\begin{equation}\label{FWH8}
\begin{aligned}
H^{\prime \prime}_{F}=&\frac{\pi^2}{2}+A^0
-\frac{\pi^4}{8}-\frac{i}{8}[\tilde{\pi}, \tilde{E}]
+\frac{\pi^6}{16}-\frac{i}{128}\left[\tilde{\pi}^{2},\{\tilde{\pi}, \tilde{E}\}\right]
+\frac{3 i}{32}\left[\tilde{\pi}^{3}, \tilde{E}\right]
+\frac{1}{16}\{\tilde{E}, \tilde{E}\}\\
&-\frac{5\pi^8}{128}+\frac{11 i}{1024}\left[\tilde{\pi}^{4},\{\tilde{\pi}, \tilde{E}\}\right]
-\frac{31 i}{512}\left[\tilde{\pi}^{5}, \tilde{E}\right]
-\frac{9 i}{512}\left[\tilde{\pi}^{3}, \tilde{\pi}^{2} \tilde{E}+\tilde{E} \tilde{\pi}^{2}
+\tilde{\pi} \tilde{E} \tilde{\pi}\right]\\
&+\frac{i}{32}\left[\partial_{t} \tilde{E}, \tilde{E}\right]
-\frac{1}{32} \beta\left\{\tilde{E},\left(2 \tilde{\pi}^{2} \tilde{E}
+2 \tilde{E} \tilde{\pi}^{2}
+\tilde{\pi} \tilde{E} \tilde{\pi}\right)\right\}
+o\left(m \alpha^{8}\right),
\end{aligned}\end{equation}
where the tilde notation is defined as $\tilde{F}=\vec{\sigma} \cdot \vec{F} $ and the mass $m$ is omitted in the expressions. The third and fourth terms in the first row correspond to the $m\alpha^4$-order corrections, the subsequent four terms in the first row belong to the $m\alpha^6$ order, and the terms in the second and third rows contribute to the $m\alpha^8$ order. When evaluating higher-order energy corrections, the number of Hamiltonian terms increases significantly. A further challenge arises from the growing severity of operator singularities at higher orders. Since the FW transformation is constructed based on the nonrelativistic expansion with respect to electron momentum and potential, both the quantity and singularity of the resulting operators increase systematically with the expansion order \citep{PhysRevA.74.022512}. Consequently, explicit regularization procedures are essential to eliminate ultraviolet divergences and yield finite, numerically computable operator expressions. Correspondingly, the difficulty of theoretical derivation rises substantially. In particular, at the $m\alpha^8$ order, complete separation of divergent high-order relativistic correction terms of hydrogen cannot be achieved, and finite physical results can only be obtained by evaluating all perturbations originating from identical physical mechanisms \citep{JPhyA.56.4}.

To resolve the aforementioned difficulties, the nonrelativistic expansion method\citep{Rutkowski_1986,Kutzelnigg1989} has been proposed in Ref.\citep{PhysRevA.111.022814}. This method decomposes the relativistic Dirac wavefunction into positive-energy and negative-energy components, expands the Dirac–Coulomb energy using $\alpha^2$ as the perturbation parameter, and establishes iterative formulas for arbitrarily high-order Dirac–Coulomb energies and wavefunctions. Nevertheless, this approach has so far only been applied to hydrogen and helium atomic systems with Slater-type basis sets, and few cross-check results are available. Furthermore, the effective operator for the Dirac–Coulomb energy at order $m\alpha^8$ has never been derived. This work focuses on effective operators of Dirac–Coulomb energies and first-order relativistic Breit potential corrections, and extends numerical calculations to two-center systems. Section II presents the theoretical framework of the nonrelativistic expansion method for two-body systems including first-order relativistic Breit potential corrections. The effective operators for the $m\alpha^8$-order effective Dirac-Coulomb energy of a two-body system in an external field and the $m\alpha^6$-order effective operator for the energy correction arising from the relativistic Breit potential are given. Although the effective Hamiltonians at orders $m\alpha^4$ and $m\alpha^6$ are finite, the singularities of the effective Hamiltonian at order $m\alpha^8$ cannot be fully separated, and the divergences must be canceled by joint computation with those arising from the second- and third-order perturbations of lower-order operators. In Section III, Gaussian basis sets are adopted to calculate the $m\alpha^8$-order ground-state energy corrections for the hydrogen atom, hydrogen molecular ion, helium atom, and hydrogen molecule, and the results are systematically compared with existing literature values. Finally, conclusions are given in the last section.

\section{NONRELATIVISTIC EXPANSION METHOD FOR TWO-BODY SYSTEMS IN EXTERNAL FIELD}
The nonrelativistic expansion method relies on the Dirac–Coulomb equation to expand Dirac wavefunctions and energies. As the leading-order term of the expansion reduces to the Schrödinger–Coulomb equation, the high-order wavefunctions from this expansion retain complete information of unperturbed Coulomb bound states. Substituting the expanded wavefunctions into the first-order relativistic Breit potential corrections yields arbitrary-order relativistic corrections for transverse one-photon-exchange interaction under the non-retardation approximation. The main method can be found in Ref.\citep{PhysRevA.111.022814}. This section mainly focuses on high-order results for two-body systems.

The two-body Dirac-Coulomb equation can be decomposed into the following coupled equations \citep{PhysRevA.111.022814}:

\begin{equation}  
\left\{\begin{aligned}  
\tilde{p}_a \Phi_a=&(E-V-M)\Phi  \\  
\tilde{p}_1 \Phi+ \tilde{p}_2 \Phi_{12}-2m_1 \Phi_1 =&(E-V-M)\Phi_1   \\
\tilde{p}_2 \Phi+ \tilde{p}_1 \Phi_{12}-2m_2 \Phi_2 =&(E-V-M)\Phi_2   \\
\tilde{p}_1 \Phi_2+ \tilde{p}_2 \Phi_1-(2m_1+2m_2) \Phi_{12} =&(E-V-M)\Phi_{12}   \\
\end{aligned}\right.  
\end{equation} 
The energy and wave function are expanded in the following manner:
\begin{equation}
\left\{\begin{aligned}
E_{DC}&=\left(m_1+m_2 \right)+\sum_i E^{(2 i)}_{DC}\\
\Phi & =\sum_{i=1}^n \Phi^{(2 i)} \\
\Phi_a & =\sum_{i=1} \Phi_a^{(2 i+1)} \\
\Phi_{12} & =\sum_{i=1} \Phi_{12}^{(2 i+2)} \\
\end{aligned}\right.
\end{equation}
By incorporating this result into the Dirac equation, we can obtain the equation that needs to be satisfied by the higher-order energy and wave function.

The leading-order wavefunctions satisfy the following system of equations:
\begin{equation}
\left\{\begin{aligned}
& H_{0} \Phi^{(2)}=E^{(2)}_{DC} \Phi^{(2)} \\
& \Phi_a^{(3)}=\frac{1}{2 m_a} \tilde{p}_a \Phi^{(2)} \\
& \Phi_{12}^{(4)}=\frac{1}{2 m_1} \frac{1}{2 m_2} \tilde{p}_1 \tilde{p}_2 \Phi^{(2)} 
\end{aligned}\right.
\end{equation}
The first line is the two-body Schrödinger equation, where $H_{0}=\sum_{i}\frac{1}{2m_i}p^2_i+V$.  The simplified recurrence relations of higher-order energy and wave functions are:
\begin{equation}
\left\{\begin{aligned}
E^{(2 n)}_{DC}&=( \frac{1}{2 m_1}+\frac{1}{2 m_2})\left(\Phi^{(2)}, \tilde{p}_1 \tilde{p}_2 \Phi_{12}^{(2 n)}\right)-\sum_a \frac{1}{2 m_a}\left(\tilde{p}_a \Phi^{(2)}, A \Phi_a^{(2 n-1)}+\sum_{i=1}^{n-2} E^{(2 i+2)}_{DC} \Phi_a^{(2 n-2 i-1)}\right), 
\\
\Phi^{(2 n)}&= G (\frac{1}{2 m_1}+\frac{1}{2 m_2}) 
\tilde{p}_1 \tilde{p}_2\Phi_{12}^{(2 n)}-\sum_a G \frac{\tilde{p}_a}{2 m_a} A \Phi_a^{(2 n-1)}-\sum_{i=1}^{n-2} E^{(2 i+2)}_{DC} G\left(\Phi^{(2 n-2 i)}+\sum_a \frac{\tilde{p}_a}{2 m_a} \Phi_a^{(2 n-2 i-1)}\right), 
\\
\Phi_1^{(2 n+1)}&=\frac{1}{2 m_1}\left(\tilde{p}_1 \Phi^{(2 n)}+\tilde{p}_2 \Phi_{12}^{(2 n)}-A \Phi_1^{(2 n-1)}-\sum_{i=1}^{n-2} E^{(2 i+2)}_{DC} \Phi_1^{(2 n-2 i-1)}\right),
\\
\Phi_2^{(2 n+1)}&=\frac{1}{2 m_2}\left(\tilde{p}_2 \Phi^{(2 n)}+\tilde{p}_1 \Phi_{12}^{(2 n)}-A \Phi_2^{(2 n-1)}-\sum_{i=1}^{n-2} E^{(2 i+2)}_{DC} \Phi_2^{(2 n-2 i-1)}\right),
\\
\Phi_{12}^{(2 n)}&=\frac{1}{2\left(m_1+m_2\right)}(\tilde{p}_1 \Phi_2^{(2 n-1)}+\tilde{p}_2 \Phi_1^{(2 n-1)}-A \Phi_{12}^{(2 n-2)}-\sum_{i=1}^{n-3} E^{(2 i+2)}_{DC} \Phi_{12}^{(2 n-2 i-2)})
\end{aligned}\right.
\end{equation}
where $A=E^{(2)}_{DC}-V$ and the nonrelativistic reduced Green operator is $G=\frac{Q}{E^{(2)}_{DC}-H_{0}}$.

From these recurrence relations, one can systematically derive corrections to the wave function and Dirac-Coulomb energy at any desired order. The regularized Hamiltonian for the Dirac-Coulomb system up to the $m\alpha^6$ order is presented in Ref.\cite{JPhyA.56.4}, and   the regularized Hamiltonian operator of leading-order relativistic energy correction to Dirac-Coulomb energy is (at $\alpha^4$ order),
\begin{equation}
H^{(4)}_{RDC}=
\frac{2m_1+2m_2}{(2m_1)^2(2m_2)^2}\tilde{p}_1^2\tilde{p}_2^2
-\frac1{(2m_1)^2}\tilde{p}_1 A \tilde{p}_1-\frac1{(2m_2)^2}\tilde{p}_2 A \tilde{p}_2,
\end{equation}
and the $\alpha^6$ order DC energy correction is 
\begin{equation}
E^{(6)}_{DC}=\langle H_{RDC}^{(6)}+H_{RDC}^{(4)}GH_{RDC}^{(4)} \rangle,
\end{equation} 
where $H_R^{(6)}$ is the $\alpha^6$ order regularized DC Hamiltonian
\begin{equation}\begin{aligned}  
H^{(6)}_{RDC}=&
\frac{\tilde{p}_1 A^{2} \tilde{p}_1}{(2m_1)^3}
+\frac{\tilde{p}_2 A^{2} \tilde{p}_2}{(2m_2)^3}
-\frac{\{\tilde{p}_1 A \tilde{p}_1, p_{2}^{2}\}}{(2m_1)^3 (2m_2)}
-\frac{\{\tilde{p}_2 A \tilde{p}_2, p_{1}^{2}\}}{(2m_1) (2m_2)^3}
-\frac{\tilde{p}_1 \tilde{p}_2 A \tilde{p}_2 \tilde{p}_1}{(2m_1)^2 (2m_2)^2}
\\&
+\dfrac{(1+m_1/m_2)p_{1}^{4}p_{2}^{2}+(1+m_2/m_1)p_{1}^{2}p_{2}^{4}}{(2m_1)^2 (2m_2)^2 (2m_1+2m_2)}
-E^{(4)}_{DC} (\dfrac{p_{1}^{2}}{(2m_{1})^{2}}+\dfrac{p_{2}^{2}}{(2m_{2})^{2}})
\end{aligned}\end{equation} 
This work presents the effective Hamiltonian up to the $m\alpha^8$ order derived from the recurrence relations.
At the $\alpha^8$ order, DC energy correction is 
\begin{equation}
E^{(8)}_{DC}=\langle H_{RDC}^{(8)}+H_{RDC}^{(6)}GH_{RDC}^{(4)}+H_{RDC}^{(4)}GH_{RDC}^{(6)} +H_{RDC}^{(4)}G(H_{RDC}^{(4)} -E^{(4)}_{DC} )GH_{RDC}^{(4)}  \rangle.
\end{equation}  
However, a crucial distinction arises. The corresponding effective Hamiltonian $H_{RDC}^{(8)}$, presented below, contains terms whose divergences cannot be separated and canceled independently of lower-order contributions.

\begin{equation}\begin{aligned}  
H^{(8)}_{RDC}=
&-\frac{\tilde{p}_1 A^3 \tilde{p}_1}{(2m_1)^4} 
- \frac{\tilde{p}_2 A^3 \tilde{p}_2}{(2m_2)^4}
+\frac{\tilde{p}_2 \tilde{p}_1 A\tilde{p}_1 A\tilde{p}_2 + \tilde{p}_2 A\tilde{p}_1 A\tilde{p}_1 \tilde{p}_2}{(2m_1)(2m_2)^3(2m_1+2m_2)}
+\frac{\tilde{p}_1 \tilde{p}_2 A\tilde{p}_2 A\tilde{p}_1 + \tilde{p}_1 A\tilde{p}_2 A\tilde{p}_2 \tilde{p}_1}{(2m_1)^3(2m_2)(2m_1+2m_2)} \\
&+\frac{\tilde{p}_1 A\tilde{p}_1 \tilde{p}_2 A\tilde{p}_2 + \tilde{p}_2 A\tilde{p}_2 \tilde{p}_1 A\tilde{p}_1}{(2m_1)^2(2m_2)^2(2m_1+2m_2)}
+\frac{\tilde{p}_1 A\tilde{p}_2^2 A\tilde{p}_1}{(2m_1)^4(2m_1+2m_2)}
+\frac{\tilde{p}_2 A\tilde{p}_1^2 A\tilde{p}_2}{(2m_2)^4(2m_1+2m_2)} \\
&-\frac{\{p_1^2,\tilde{p}_2 A^2 \tilde{p}_2\}}{(2m_1)(2m_2)^4}
-\frac{\{p_2^2,\tilde{p}_1 A^2 \tilde{p}_1\}}{(2m_1)^4(2m_2)}
+\frac{\tilde{p}_1 \tilde{p}_2 A^2 \tilde{p}_1 \tilde{p}_2}{(2m_1)^2(2m_2)^2(2m_1+2m_2)} \\
&-\frac{\{p_1^2 p_2^2,\tilde{p}_2 A\tilde{p}_2\}+\{p_1^2,\tilde{p}_1 \tilde{p}_2 A\tilde{p}_1 \tilde{p}_2\}}{(2m_1)^2(2m_2)^3(2m_1+2m_2)}
-\frac{\{p_1^2 p_2^2,\tilde{p}_1 A\tilde{p}_1\}+\{p_2^2,\tilde{p}_1 \tilde{p}_2 A\tilde{p}_1 \tilde{p}_2\}}{(2m_1)^3(2m_2)^2(2m_1+2m_2)} 
\\&
-\frac{\{p_1^4,\tilde{p}_2 A\tilde{p}_2\}}{(2m_1)(2m_2)^4(2m_1+2m_2)}
-\frac{\{p_2^4,\tilde{p}_1 A\tilde{p}_1\}}{(2m_1)(2m_2)^4(2m_1+2m_2)}
\\&
-\frac{p_2^2 \tilde{p}_1 A\tilde{p}_1 p_2^2}{(2m_1)^4(2m_2)^2}
-\frac{p_1^2 \tilde{p}_2 A\tilde{p}_2 p_1^2}{(2m_1)^2(2m_2)^4} 
+\frac{(m_2/m_1)p_1^2 p_2^6+(m_1/m_2)p_2^2 p_1^6+2p_2^4 p_1^4}{(2m_1)^3(2m_2)^3(2m_1+2m_2)} 
\\&
+\left(\frac{2\tilde{p}_1 A\tilde{p}_1}{(2m_1)^3}
+\frac{2\tilde{p}_2 A\tilde{p}_2}{(2m_2)^3}
-\frac{2m_1/m_2+2m_2/m_1+1}{(2m_1)^2(2m_2)^2}p_1^2 p_2^2
\right)E^{(4)}_{DC}
-\frac{E^{(6)}_{DC} p_2^2}{(2m_2)^2}-\frac{E^{(6)}_{DC} p_1^2}{(2m_1)^2}.
\end{aligned}\end{equation}

Although we have derived explicit expressions for $H^{(4)}_{RDC},H^{(6)}_{RDC}$ and $H^{(8)}_{RDC}$, direct numerical evaluation of the energy at $m\alpha^8$ using these operators alone is not feasible. The singularities embedded in $H^{(8)}_{RDC}$—evident from the high powers of momentum and potential—cannot be regularized independently; they must cancel against divergences arising from second- and third-order perturbations of lower-order Hamiltonians. This point becomes especially evident in one-body systems \cite{JPhyA.56.4}: $H^{(4)}_{RDC}$ and $H^{(6)}_{RDC}$ are already the simplest finite regularized operators, yet their second- and third-order perturbative corrections remain divergent; these divergences must be canceled against those of $H^{(8)}_{RDC}$ to yield a finite $E^{(8)}_{DC}$. Hence one needs to incorporate $H^{(8)}_{RDC}$ into the higher-order perturbation expansion of the lower-order Hamiltonians—a procedure equivalent to the iterative expansion method introduced above \cite{JPhyA.56.4}.

In the limit $m_2\gg m_1$ (the single-electron approximation), the Hamiltonians at each order are:

\begin{equation}
H^{(4)}_{RDC}=
-\frac1{(2m_1)^2}\tilde{p}_1 A \tilde{p}_1+\frac{1}{(2m_1)^2(2m_2)}\tilde{p}_1^2\tilde{p}_2^2+o(m_2^{-1}),
\end{equation}

\begin{equation}\begin{aligned}  
H^{(6)}_{RDC}=
\frac{\tilde{p}_1 A^{2} \tilde{p}_1}{(2m_1)^3}
-\frac{\{\tilde{p}_1 A \tilde{p}_1, p_{2}^{2}\}}{(2m_1)^3 (2m_2)}
- \dfrac{E^{(4)}_{DC} p_{1}^{2}}{(2m_{1})^{2}}+o(m_2^{-1}),
\end{aligned}\end{equation}

\begin{equation}\begin{aligned}  
H^{(8)}_{RDC}=
&-\frac{\tilde{p}_1 A^3 \tilde{p}_1}{(2m_1)^4} 
+\frac{2\tilde{p}_1 A\tilde{p}_1}{(2m_1)^3}E^{(4)}_{DC}
-\frac{E^{(6)}_{DC} p_1^2}{(2m_1)^2}
\\&
+\frac{\tilde{p}_1 A\tilde{p}_2^2 A\tilde{p}_1}{(2m_1)^4(2m_2)} 
-\frac{\{p_2^2,\tilde{p}_1 A^2 \tilde{p}_1\}}{(2m_1)^4(2m_2)} 
-\frac{2p_1^2 p_2^2E^{(4)}_{DC}}{(2m_1)^3(2m_2)}+o(m_2^{-1}).
\end{aligned}\end{equation}
The first terms of $H^{(4)}_{RDC}$ and $H^{(6)}_{RDC}$, as well as the first three terms of $H^{(8)}_{RDC}$, agree with the single-electron results; the remaining terms are recoil corrections. In the $m_2\gg m_1$ limit, divergences within each order must be canceled by their respective recoil contributions, and the leading-order parts reduce to the single-electron expressions.  It is readily seen that in the no-recoil approximation, the expectation values of $H^{(4)}_{RDC}$ and $H^{(6)}_{RDC}$ are finite, but their perturbative expansions diverge, so their divergences must cancel against those from $H^{(8)}_{RDC}$. Meanwhile, at the first-order recoil level, the last three (recoil) terms in $H^{(8)}_{RDC}$ also carry singularities, which must be combined with lower-order recoil contributions to yield a finite result. Thus nonseparable divergences appear ubiquitously at third order of relativistic corrections. The iterative expansion method circumvents the appearance of these divergences, and numerical calculations (see Sec. III) confirm its validity.

The relativistic correction defined in this work is
\begin{equation}
E_{\text{rel}}=E_{\text{DC}}+\frac{\langle\Psi|V_T|\Psi\rangle}{\langle\Psi|\Psi\rangle}.
\end{equation}
where the second term corresponds to the relativistic Breit potential 
$
V_T=\frac{q_{1}q_{2}}{4\pi r_{12}}
\left(\boldsymbol{\alpha}_1 \cdot \boldsymbol{\alpha}_2+\frac{\left(\boldsymbol{\alpha}_1 \cdot \boldsymbol{r}_{12}\right)\left(\boldsymbol{\alpha}_2 \cdot \boldsymbol{r}_{12}\right)}
{r_{12}^2}\right).
$
This term arises from the transverse one-photon-exchange interaction under the non-retardation approximation.

The relativistic Breit potential corrections can be expanded by means of iterative relations.
\begin{equation}
\begin{aligned}
E_{Breit}^{(4)}=&(\langle\Psi|V_T|\Psi\rangle)^{(4)},
\\
E_{Breit}^{(6)}=&(\langle\Psi|V_T|\Psi\rangle)^{(6)}
-(\langle\Psi|V_T|\Psi\rangle)^{(4)}\langle\Psi|\Psi\rangle^{(2)},
\\
E_{Breit}^{(8)}=&(\langle\Psi|V_T|\Psi\rangle)^{(8)}
-(\langle\Psi|V_T|\Psi\rangle)^{(6)}\langle\Psi|\Psi\rangle^{(2)}
-(\langle\Psi|V_T|\Psi\rangle)^{(4)}
(\langle\Psi|\Psi\rangle^{(4)}+(\langle\Psi|\Psi\rangle^{(2)})^2).
\end{aligned}
\end{equation}
where
\begin{equation}
\begin{aligned}
\langle\Psi|\Psi\rangle^{(2n)}=
& \sum_{i=0}^{n}
\langle\Phi^{(2+2i)}|\Phi^{(2+2n-2i)}\rangle
+\sum_{a}\sum_{i=0}^{n-1}
\langle\Phi_a^{(3+2i)}| \Phi_a^{(1+2n-2i)}\rangle
+\sum_{i=0}^{n-2}
\langle\Phi_{12}^{(4+2i)}| \Phi_{12}^{(2n-2i)}\rangle,
\end{aligned}
\end{equation}
\begin{equation}
\begin{aligned}
(\langle\Psi|V_T|\Psi\rangle)^{(4+2n)}=
& \sum_{i=0}^{n}\left(
\langle\Phi^{(2+2i)}| V^t |\Phi_{12}^{(4+2n-2i)}\rangle
+\langle\Phi_{1}^{(3+2i)}| V^t |\Phi_{2}^{(3+2n-2i)}\rangle
\right.\\&
\left.
+\langle\Phi_{2}^{(3+2i)}| V^t |\Phi_{1}^{(3+2n-2i)}\rangle
+\langle\Phi_{12}^{(4+2i)}| V^t |\Phi^{(2+2n-2i)}\rangle\right).
\end{aligned}
\end{equation}
and $V^t=\frac{q_{1}q_{2}}{4\pi r_{12}}\left(\sigma_1 \cdot \sigma_2-\frac{\left(\sigma_1 \cdot r_{12}\right)\left(\sigma_2 \cdot r_{12}\right)}{r_{12}^2}\right)$.

The leading term is given by
\begin{equation}
\begin{aligned}
E_{Breit}^{(4)}=(\langle\Psi|V_T|\Psi\rangle)^{(4)}_{2-body}
=\left\langle \Phi^{(2)} | V^{(4)}_{1p} | \Phi^{(2)} \right\rangle
\end{aligned}
\end{equation}
and the nonrelativistic potential of transverse single-photon exchange interaction is
$V^{(4)}_{1p}=  \left\{ \frac{\tilde{p}_1}{m_1},\left\{ \frac{\tilde{p}_2}{m_2}, V^t \right\} \right\}$. 
Higher-order terms correspond to mixed perturbative corrections arising from first-order transverse single-photon exchange and Coulomb photon exchange interactions.
\begin{equation}
\begin{aligned}
(\langle\Psi|V_T|\Psi\rangle)^{(6)}_{2-body}=
\langle\Phi^{(2)}|V^{(4)}_{1p}G H_{RDC}^{(4)} |\Phi^{(2)}\rangle
+\langle\Phi^{(2)}|H_{RDC}^{(4)} G V^{(4)}_{1p}|\Phi^{(2)}\rangle
+\langle\Phi^{(2)}| V^{(6)}_{1p}|\Phi^{(2)}\rangle
\end{aligned}
\end{equation}
The first two terms are mixed perturbative corrections from transverse single-photon exchange and Coulomb photon exchange interactions, while the third term represents high-order relativistic corrections originating from the high-order Breit potential
\begin{equation}
\begin{aligned}
V^{(6)}_{1p}&= \frac{\{\widetilde{p}_1\widetilde{p}_2,\,(\widetilde{p}_2 V^t \widetilde{p}_2 + \widetilde{p}_1 V^t \widetilde{p}_1)\}}{(2m_1)^2(2m_2)^2}
+ \frac{1}{(2m_1+2m_2)} \frac{1}{2m_1} \frac{1}{2m_2} \left\{ \left( \frac{1}{2m_2}\widetilde{p}_1^3 \widetilde{p}_2 + \frac{1}{2m_1}\widetilde{p}_1 \widetilde{p}_2^3 \right), V^t \right\} \\
&\quad - \frac{1}{(2m_1)^2} \frac{1}{2m_2} \left( \widetilde{p}_1 A V^t \widetilde{p}_2 + \widetilde{p}_2 V^t A \widetilde{p}_1 \right)
- \frac{1}{(2m_2)^2} \frac{1}{2m_1} \left( \widetilde{p}_2 A V^t \widetilde{p}_1 + \widetilde{p}_1 V^t A \widetilde{p}_2 \right) \\
&\quad - \frac{1}{(2m_1+2m_2)} \frac{1}{(2m_2)^2} \left( \widetilde{p}_2 A \widetilde{p}_1 V^t + V^t \widetilde{p}_1 A \widetilde{p}_2 \right)
- \frac{1}{(2m_1+2m_2)} \frac{1}{(2m_1)^2} \left( V^t \widetilde{p}_2 A \widetilde{p}_1 + V^t \widetilde{p}_2 A \widetilde{p}_1 \right) \\
&\quad - \frac{1}{(2m_1+2m_2)} \frac{1}{2m_1} \frac{1}{2m_2} \left( \widetilde{p}_1 \widetilde{p}_2 A V^t + V^t A \widetilde{p}_1 \widetilde{p}_2 \right)
\end{aligned}
\end{equation}
It can be seen that the singularity of the operator intensifies as the order increases, and divergences that cannot be fully separated are very likely to exist in the effective operator at the $m\alpha^8$ order. Nevertheless, the iterative expansion method yields finite results even at the $m\alpha^8$ order. Furthermore, second- and higher-order perturbative corrections involving only transverse photons are neglected. A consistent evaluation of such terms would require inclusion of all two-photon-exchange diagrams within the nonrelativistic QED framework, which is beyond the scope of this work. Therefore, we restrict our calculation to first order in the transverse photon interaction and only include its mixing with Coulomb interactions; complete two-photon effects will be addressed in future studies.

\section{CALCULATION METHOD AND NUMERICAL RESULTS}

In this section, the nonrelativistic expansion method is employed to calculate high-order relativistic corrections to the ground-state energies of atomic hydrogen, the hydrogen molecular ion, helium atom, and hydrogen molecule. The Born-Oppenheimer approximation is adopted for two-center systems. Two types of single-electron Gaussian basis sets, denoted as $ e^{-\alpha^2 r^2}\chi_s$ and $\boldsymbol{\sigma}\cdot\mathbf{r} e^{-\alpha^2 r^2}\chi_s$, are used in the ground-state calculation. This is because in the nonrelativistic expansion method, we need to compute the matrix of the operator $\boldsymbol{\sigma}\cdot\mathbf{p}$, which flips the quantum number $\kappa$ of the quantum state to $-\kappa$. Introducing the $\boldsymbol{\sigma}\cdot\mathbf{r}$ term in the Gaussian basis sets achieves the same effect on the quantum number $\kappa$ and ensures that the matrix of the operator $\boldsymbol{\sigma}\cdot\mathbf{p}$ is non-zero. Furthermore, one of these two Gaussian basis sets consists solely of s-wave functions, and the other employs p-wave functions. Compared with conventional Gaussian basis sets, this choice reduces the number of p-wave basis functions and saves computational resources. For two-electron systems, we take the direct product of single-electron Gaussian basis sets, so that a two-electron system has four Gaussian basis sets: $ e^{-\alpha_a^2 r^2_a-\alpha_b^2 r^2_b}\chi_{s_a}\chi_{s_b}$ , $ \boldsymbol{\sigma}_a\cdot\mathbf{r}_a e^{-\alpha_a^2 r^2_a-\alpha_b^2 r^2_b}\chi_{s_a}\chi_{s_b}$, $ \boldsymbol{\sigma}_b\cdot\mathbf{r}_b e^{-\alpha_a^2 r^2_a-\alpha_b^2 r^2_b}\chi_{s_a}\chi_{s_b}$, and $ \boldsymbol{\sigma}_a\cdot\mathbf{r}_a \boldsymbol{\sigma}_b\cdot\mathbf{r}_b e^{-\alpha_a^2 r^2_a-\alpha_b^2 r^2_b}\chi_{s_a}\chi_{s_b}$.

To benchmark our implementation against a known analytical solution, we first present results for atomic hydrogen. Tables I and II summarize these calculations using Gaussian basis sets of increasing size. The calculated high-order energy corrections are free from divergences. The high-order energy corrections can converge to exact values with a small number of Gaussian basis functions. Calculations with 48 basis functions converge to 3–4 significant digits within order $\alpha^{10}$, while results from order $\alpha^{12}$ to $\alpha^{20}$ converge to 1–2 significant digits. Highly accurate numerical results of relevant calculations have been provided with Slater basis sets in Ref. \citep{PhysRevA.111.022814}.

We next examine a two-center system—the hydrogen molecular ion at its equilibrium geometry ($R = 2.0 \ \ a.u.$). Table III presents results up to order $\alpha^{10}$ using basis functions that include only s- and p-wave components. No divergence appears in the numerically evaluated high-order energy corrections for this two-center system, yet the converged values deviate from those reported in Ref. \citep{PhysRevA.105.L060801}. This discrepancy arises because only s- and p-wave basis functions are adopted within the Gaussian basis set used for Table III. The limited angular flexibility of an s/p-only basis introduces noticeable errors in higher-order corrections. To address this, we augment our basis set with d-wave functions. As shown in Table IV, this significantly improves agreement with high-precision reference data \citep{PhysRevA.105.L060801}, validating the accuracy and convergence behavior of our approach for two-center systems. We further report results up to order $\alpha^{10}$.

For a few-electron benchmark, we consider the helium atom ground state. Table V presents our Dirac-Coulomb energy corrections up to  $\alpha^{8}$ order alongside reference values obtained with Slater-type basis sets \cite{PhysRevA.111.022814}. The agreement between these two independent approaches confirms the reliability of both calculations. The operator expectation values required for the expansion of high-order relativistic corrections arising from the first-order relativistic Breit potential corrections of the helium ground state are listed in Table VI. $\langle\Psi|V_T|\Psi\rangle^{(4)}$ denotes the leading-order correction from the Breit potential, and $\langle\Psi|\Psi\rangle^{(4)}$ corresponds to the normalization factor of the Dirac wave function. The relation $\langle\Psi|\Psi\rangle^{(4)}=\langle\sum_a \frac{p_a^2}{(2m)^2}\rangle=\frac{1}{2m}E_{DC}^{(2)}$ implies that $\langle\Psi|\Psi\rangle^{(4)}$ equals half of the nonrelativistic ionization energy, and the numerical results also basically satisfy this equality. No numerical divergences are encountered throughout all calculations. Table VII presents the high-order relativistic corrections originating from the Breit potential. The results converge to 3–4 significant digits for terms up to order $\alpha^{6}$. The order-$\alpha^{8}$ correction has not yet converged to a reliable value, although a clear convergence trend is observed, implying that more accurate results can be achieved by further enlarging the basis set. In addition, the estimated magnitude of this order correction is comparable to the numerical uncertainty of the Dirac-Coulomb energy; therefore, we omit this correction in our final results.

Our final benchmark is the ground state of hydrogen molecule at its equilibrium internuclear separation ($R = 1.4 \ \  a.u.$). Basis functions are constructed as direct products of one-electron Gaussian orbitals centered on each nucleus, retaining only s- and p-wave components. Based on our experience with the hydrogen molecular ion, where a similar limitation led to errors of about $5\%$ in higher-order corrections, we adopt a conservative error estimate of  $10\%$ for the present hydrogen molecule results (Table VIII).

\begin{table}[htbp]
\caption{High-order relativistic energy corrections for the ground state of the hydrogen atom calculated via the nonrelativistic expansion method with Gaussian basis sets. The first column lists the number of basis functions, the second column shows the nonrelativistic energy, while subsequent columns present the relativistic high-order corrections $E^{(2n)}$ divided by $\alpha^{2n}$ in atomic units up to order $\alpha^{10}$.}
\begin{tabular}{llllll}
\hline
$N$ & $E^{(2)}_{DC}$ & $E^{(4)}_{DC}$ & $E^{(6)}_{DC}$ & $E^{(8)}_{DC}$ & $E^{(10)}_{DC}$\\
\hline\hline
12 & -0.480349 & -0.064862 & 0.000952 & 0.002926 & -0.0004 \\[3pt]
16 & -0.491950 & -0.090499 & -0.01170 & 0.002598 & 0.0015 \\[3pt]
20 & -0.495663 & -0.096021 & -0.01767 & 0.001677 & 0.0020 \\[3pt]
24 & -0.499079 & -0.114352 & -0.03929 & -0.00990 & 0.0001 \\[3pt]
28 & -0.499912 & -0.123155 & -0.05672 & -0.02870 & -0.0136 \\[3pt]
32 & -0.499983 & -0.124354 & -0.06008 & -0.03399 & -0.0196 \\[3pt]
36 & -0.499997 & -0.124824 & -0.06172 & -0.03718 & -0.0240 \\[3pt]
40 & -0.499999 & -0.124934 & -0.06218 & -0.03819 & -0.0256 \\[3pt]
44 & -0.500000 & -0.124979 & -0.06238 & -0.03871 & -0.0266 \\[3pt]
48 & -0.500000 & -0.124993 & -0.06246 & -0.03893 & -0.0270 \\[3pt]
$\infty$ & -0.500000 & -0.125000 (7) & -0.06251 (5) & -0.03909 (16) & -0.0274 (4) \\[3pt]
Exact & -0.5 & -0.125 & -0.0625 & -0.0390625 & -0.0273... \\[3pt]
\hline
\end{tabular}
\end{table}

\begin{table}[htbp]
\caption{High-order relativistic energy corrections for the ground state of the hydrogen atom calculated via the nonrelativistic expansion method with Gaussian basis sets. The first column lists the number of basis functions,  while subsequent columns present the relativistic corrections $E^{(2n)}$ divided by $\alpha^{2n}$ in atomic units from order $\alpha^{12}$ to $\alpha^{20}$.}
\begin{tabular}{llllll}
\hline
$N$ & $E^{(12)}_{DC}$ & $E^{(14)}_{DC}$ & $E^{(16)}_{DC}$ & $E^{(18)}_{DC}$ & $E^{(20)}_{DC}$\\
\hline\hline
12 & -0.0002 & 0.0001 & 0.0000 & -0.0000 &  0.0000 \\[3pt]
16 & -0.0001 & -0.0002 & -0.0000 & 0.0000 &  0.0000 \\[3pt]
20 & 0.0002 & -0.0003 & -0.0001 & 0.0000 & 0.0000 \\[3pt]
24 & 0.0016 & 0.0008 & 0.0001 & -0.0001 & -0.0001 \\[3pt]
28 & -0.0054 & -0.0014 & 0.0002 & 0.0006 & 0.0004 \\[3pt]
32 & -0.0108 & -0.0054 & -0.0028 & -0.0007 & 0.0001 \\[3pt]
36 & -0.0157 & -0.0101 & -0.0062 & -0.0036 & -0.0020 \\[3pt]
40 & -0.0178 & -0.0125 & -0.0086 & -0.0058 & -0.0038 \\[3pt]
44 & -0.0192 & -0.0142 & -0.0105 & -0.0078 & -0.0057 \\[3pt]
48 & -0.0199 & -0.0152 & -0.0118 & -0.0092 & -0.0072 \\[3pt]
$\infty$ & -0.0207 (8) & -0.0166 (14) & -0.0139 (21) & -0.0125 (33) & -0.012 (5) \\[3pt]
Exact & -0.0205... & -0.0161... & -0.0130... & -0.0109... & -0.0092... \\[3pt]
\hline
\end{tabular}
\end{table}

\begin{table}[htbp]
\caption{High-order energy corrections for the hydrogen molecular ion at $R=2$, computed using the nonrelativistic expansion method. Only s-wave and p-wave basis functions are employed in the Gaussian basis set. The last row presents the results taken from Ref.\cite{PhysRevA.105.L060801}. It can be observed that due to the limited number of partial waves included, the present results only achieve one or two significant digits. This work reports the relativistic energy corrections for the ground state of the hydrogen molecular ion up to order $\alpha^{10}$. The first column denotes the number of basis functions, while the remaining columns correspond to $E^{(2n)}$ divided by $\alpha^{2n}$ in atomic units.}
\begin{tabular}{llllll}
\hline
$N$ & $E^{(2)}_{DC}$ & $E^{(4)}_{DC}$ & $E^{(6)}_{DC}$ & $E^{(8)}_{DC}$ & $E^{(10)}_{DC}$\\
\hline\hline
24 & -1.066967324 & -0.067741 & 0.0152092 & -0.00143 & -0.00035 \\[3pt]
32 & -1.083011802 & -0.088890 & 0.0025515 & 0.002184 & 0.000024 \\[3pt]
40 & -1.090052762 & -0.096731 & -0.000103 & 0.001725 & 0.000179 \\[3pt]
48 & -1.094594674 & -0.119130 & -0.019275 & -0.00433 & -0.00003 \\[3pt]
56 & -1.095707920 & -0.130128 & -0.036804 & -0.01949 & -0.00921 \\[3pt]
64 & -1.095796286 & -0.131844 & -0.041150 & -0.02552 & -0.01531 \\[3pt]
72 & -1.095809960 & -0.132312 & -0.042578 & -0.02816 & -0.01812 \\[3pt]
80 & -1.095812985 & -0.132479 & -0.043297 & -0.02948 & -0.02045 \\[3pt]
88 & -1.095813196 & -0.132510 & -0.043435 & -0.02978 & -0.02119 \\[3pt]
$\infty$ & -1.095813212(6) & -0.132517(7) & -0.043467(33) & -0.02987(9) & -0.02153(34)  \\[3pt]
Ref.\cite{PhysRevA.105.L060801} & -1.1026... & -0.138332... & -0.0417.. & -0.02831... &    \\[3pt]
\hline
\end{tabular}
\end{table}

\begin{table}[htbp]
\caption{High-order energy corrections for the hydrogen molecular ion at $R=2$, computed using the nonrelativistic expansion method. S-wave, p-wave, and d-wave basis functions are employed in the Gaussian basis set. The last row presents the results taken from Ref.\cite{PhysRevA.105.L060801}. Benefiting from the addition of d-wave partial waves, the present numerical results achieve 3–4 significant digits consistency with the reference data. This work reports the relativistic energy corrections for the ground state of the hydrogen molecular ion up to order $\alpha^{10}$. The first column denotes the number of basis functions, while the remaining columns correspond to the $E^{(2n)}$ divided by $\alpha^{2n}$ in atomic units.}
\begin{tabular}{llllll}
\hline
$N$ & $E^{(2)}_{DC}$ & $E^{(4)}_{DC}$ & $E^{(6)}_{DC}$ & $E^{(8)}_{DC}$ & $E^{(10)}_{DC}$\\
\hline\hline
120 & -1.083629982 & -0.075618 & 0.0130884 & 0.000183 & -0.000763 \\[3pt]
160 & -1.093843122 & -0.096486 & 0.0023419 & 0.002562 & 0.000041 \\[3pt]
200 & -1.098320977 & -0.103816 & -0.000062 & 0.002015 & 0.000213 \\[3pt]
240 & -1.101772636 & -0.125257 & -0.018101 & -0.003724 & -0.000077 \\[3pt]
280 & -1.102532006 & -0.135777 & -0.035204 & -0.018469 & -0.009018 \\[3pt]
320 & -1.102608817 & -0.137429 & -0.039362 & -0.024509 & -0.014937 \\[3pt]
360 & -1.102625757 & -0.137928 & -0.040766 & -0.026675 & -0.017956 \\[3pt]
400 & -1.102628635 & -0.138032 & -0.041128 & -0.027436 & -0.019149 \\[3pt]
440 & -1.102628641 & -0.138082 & -0.041409 & -0.028126 & -0.020252 \\[3pt]
480 & -1.102628743 & -0.138108 & -0.041471 & -0.028238 & -0.020493 \\[3pt]
$\infty$ & -1.10262878(5) & -0.138121(13) & -0.041503(32) & -0.02832(8) & -0.02061(12)  \\[3pt]
Ref.\cite{PhysRevA.105.L060801} & -1.102634\dots & -0.138332\dots & -0.0417\dots & -0.02831\dots &  \\[3pt]
\hline
\end{tabular}
\end{table}

\begin{table}[htbp]
\caption{The relativistic corrections to the energy of the ground state of a Helium atom up to $\alpha^{8}$. The last row presents the results taken from Ref.\cite{PhysRevA.111.022814}. The first column is the number of basis functions and the remaining columns are the relativistic higher-order corrections $E^{(2n)}_{DC}$ divided by $\alpha^{2n}$ in atomic units.}
\begin{tabular}{lllll}
\hline
$N$ & $E^{(2)}_{DC}$ & $E^{(4)}_{DC}$ & $E^{(6)}_{DC}$ & $E^{(8)}_{DC}$ \\
\hline\hline
400      & -2.846483 & -2.5152 & -4.264 & -7.43  \\[3pt]
576      & -2.881943 & -2.4964 & -4.325 & -8.26  \\[3pt]
784      & -2.885832 & -2.4917 & -4.379 & -8.67  \\[3pt]
1024     & -2.886273 & -2.4927 & -4.412 & -8.93  \\[3pt]
1296     & -2.886349 & -2.4943 & -4.437 & -9.12  \\[3pt]
1600     & -2.886380 & -2.4959 & -4.457 & -9.28  \\[3pt]
1936     & -2.886399 & -2.4972 & -4.474 & -9.41  \\[3pt]
2304     & -2.886413 & -2.4983 & -4.488 & -9.52 \\[3pt]
2704     & -2.886424 & -2.4992 & -4.500 & -9.61  \\[3pt]
$\infty$ & -2.88646(4) & -2.504(5) & -4.57(7) & -10.1(5)  \\[3pt]
ref.\citep{PhysRevA.111.022814} & -2.886486(6) & -2.5085(31)&  -4.55(11) &  -9.97(38) \\[3pt]
\hline
\end{tabular}
\end{table}

\begin{table}[htbp]
\caption{Operator expectation values for the high-order relativistic expansion of the Breit potential energy of the helium atom ground state. The table lists the $\langle\Psi|V_T|\Psi\rangle^{(4,6,8)}$ and the Dirac wave function normalization factors $\langle\Psi|\Psi\rangle^{(4,6)}$. All calculations exhibit stable numerical convergence without divergence. The first column is the number of basis functions and the remaining columns list the expectation values divided by  $\alpha^{2n}$ in atomic units.}
\begin{tabular}{llllll}
\hline
$N$ & $\langle\Psi|V_T|\Psi\rangle^{(4)}$ & $\langle\Psi|V_T|\Psi\rangle^{(6)}$ & $\langle\Psi|V_T|\Psi\rangle^{(8)}$ & $\langle\Psi|\Psi\rangle^{(4)}$ & $\langle\Psi|\Psi\rangle^{(6)}$ \\
\hline\hline
400  & 0.16123 & 0.7518 & 3.128 & 1.43398 & 4.7016 \\[3pt]
576  & 0.17839 & 0.9037 & 2.768 & 1.44216 & 4.7583 \\[3pt]
784  & 0.17621 & 0.9439 & 2.813 & 1.44270 & 4.7808 \\[3pt]
1024 & 0.17483 & 0.9628 & 2.805 & 1.44292 & 4.7892 \\[3pt]
1296 & 0.17420 & 0.9719 & 2.836 & 1.44299 & 4.7941 \\[3pt]
1600 & 0.17387 & 0.9772 & 2.852 & 1.44303 & 4.7985 \\[3pt]
1936 & 0.17368 & 0.9808 & 2.868 & 1.44307 & 4.8022 \\[3pt]
2304 & 0.17356 & 0.9833 & 2.883 & 1.44310 & 4.8053 \\[3pt]
2704 & 0.17347 & 0.9851 & 2.896 & 1.44312 & 4.8079 \\[3pt]
$\infty$& 0.17330(17) & 0.991(6) & 2.99(9) &1.44321(9)&4.823(15)\\[3pt]
\hline
\end{tabular}
\end{table}

\begin{table}[htbp]
\caption{The high-order relativistic expansion of the Breit potential energy of the helium atom ground state. The first column is the number of basis functions and the remaining columns are the relativistic higher-order corrections $E^{(2n)}_{Breit}$ divided by $\alpha^{2n}$ in atomic units.}
\begin{tabular}{llll}
\hline
$N$ & $   E^{(4)}_{Breit}$ & $E^{(6)}_{Breit}$ & $E^{(8)}_{Breit}$ \\
\hline\hline
400      & 0.16123 & 0.521 & 0.960 \\[3pt]
576      & 0.17839 & 0.646 & 0.245 \\[3pt]
784      & 0.17621 & 0.690 & 0.242 \\[3pt]
1024     & 0.17483 & 0.711 & 0.214 \\[3pt]
1296     & 0.17420 & 0.720 & 0.235 \\[3pt]
1600     & 0.17387 & 0.726 & 0.246 \\[3pt]
1936     & 0.17368 & 0.730 & 0.257 \\[3pt]
2304     & 0.17356 & 0.733 & 0.269 \\[3pt]
2704    & 0.17347 & 0.735 & 0.279 \\[3pt]
$\infty$ & 0.17330(17) & 0.740(5) & 0.37(9) \\[3pt]
\hline
\end{tabular}
\end{table}

\begin{table}[htbp]
\caption{The relativistic corrections to the energy of the ground state of the hydrogen molecule up to $\alpha^{8}$.  The first column is the number of basis functions and the remaining columns are the relativistic higher-order corrections $E^{(2n)}$ divided by $\alpha^{2n}$ in atomic units. Referring to results for the hydrogen molecular ion with only $s$- and $p$-wave partial waves retained, the nonrelativistic energy error in the last row is estimated to be $0.1\%$, while the error for high-order corrections is $10\%$.}
\begin{tabular}{llllllll}
\hline
$N$ & $E^{(2)}_{DC}$ & $E^{(4)}_{DC}$ & $E^{(6)}_{DC}$ & $E^{(8)}_{DC}$ & $E^{(4)}_{Breit}$ & $E^{(6)}_{Breit}$ & $E^{(8)}_{Breit}$  \\
\hline\hline

512 & -1.84574 & -0.282 & -0.0665 & -0.0289 & -0.0994 & -0.0343 & 0.0882 \\[3pt]
800 & -1.86679 & -0.271 & -0.0710 & -0.0376 & -0.0894 & -0.0285 & 0.0365 \\[3pt]
1152 & -1.86768 & -0.272 & -0.0732 & -0.0417 & -0.0888 & -0.0292 & 0.0183 \\[3pt]
1568 & -1.86784 & -0.272 & -0.0746 & -0.0445 & -0.0889 & -0.0294 & 0.0223 \\[3pt]
2048 & -1.86791 & -0.272 & -0.0752 & -0.0449 & -0.0889 & -0.0295 & 0.0235 \\[3pt]
$\infty$ & -1.86797(6) & -0.27(3) & -0.076(8) & -0.045(5) 
& -0.089(9) & -0.030(3) & 0.024(3) \\[3pt]
\hline
\end{tabular}
\end{table}

\section{CONCLUSIONS}

This work investigates the nonrelativistic expansion of the two-body Dirac-Coulomb system with arbitrary masses, and obtains results up to order $m\alpha^8$. By expanding the effective energy-correction operator in terms of the mass ratio $\frac{m_1}{m_2}$, we find that effective Hamiltonian operators with incompletely separable divergences, analogous to those in hydrogen-like systems, emerge at order $m\alpha^8$. Their divergences must cancel jointly with the divergences arising from higher-order perturbations of lower-order Hamiltonian operators. Numerical calculations within the nonrelativistic-expansion formalism are exactly consistent with this cancellation scheme. The singularities of the effective operators obtained from the nonrelativistic expansion of the energy-correction operator for the first order relativistic Breit potential corrections also grow with increasing order. Consequently, the energy corrections originating from the Dirac-Coulomb energy and the first order relativistic Breit potential corrections are not easily computed at high orders numerically via the conventional effective-operator approach.

In our numerical calculations of high-order relativistic corrections for the ground states of the hydrogen atom, hydrogen molecular ion, helium atom, and hydrogen molecule using Gaussian basis sets, we encounter no divergent matrix elements of intermediate operators and obtain reasonably good results. For the hydrogen atom, convergence to the exact solution up to order $m\alpha^{20}$ is readily achieved. The results for the hydrogen molecular ion are consistent with existing numerical results at order $m\alpha^8$ and extend to order $m\alpha^{10}$. The Dirac-Coulomb energies for the helium atom also agree with our earlier calculations based on Slater-type basis sets. 

A current limitation of this approach is that our numerical implementation does not yet include explicit electron-correlation effects (e.g., the use of explicitly correlated Gaussian basis sets), which limits the attainable precision for few-electron systems. Future work will aim to integrate such techniques and to extend the present framework to QED corrections at higher orders.

\textbf{ACKNOWLEDGMENTS}
This work was supported by the National Natural Science Foundation of China (Grants No. 12074295 and No. 12304271).

\bibliography{1.bib}

\end{document}